\documentclass[12pt]{iopart}
\usepackage{graphicx}
\usepackage{setspace}
\usepackage[square,sort&compress,numbers]{natbib}
\usepackage{xurl}

\begin{document}

\title{Sample thickness dependence of structural and magnetic properties in $\alpha$-RuCl$_3$}

\author{Paige Harford$^1$, Ezekiel Horsley$^1$, 
Subin Kim$^1$ and Young-June Kim$^1$}

\address{$^1$ Department of Physics, University of Toronto, Toronto, ON M5S~1A7, Canada}

\ead{youngjune.kim@utoronto.ca}

\vspace{10pt}
\begin{indented}
\item[]December 2025
\end{indented}

\begin{abstract}
The layered transition metal trihalide $\alpha$-RuCl$_3$ has been studied extensively in recent years as a promising candidate for a proximate Kitaev quantum spin liquid state. In high quality samples, a complete structural transition from room-temperature C2/m to low-temperature R$\bar{3}$ is consistently observed, with a single magnetic transition to antiferromagnetic ordering at $\sim$7K. However, magnetic and physical properties have been shown to depend heavily on both sample size and sample quality, with small and damaged samples exhibiting incomplete structural transitions and multiple magnetic anomalies. Although large high quality samples have been well studied, an understanding of the features attributed to low quality or small sample size is limited. Here, we probe the structural and magnetic transitions of $\alpha$-RuCl$_3$ single crystal samples via magnetic susceptibility through a range of thickness, manipulated through careful mechanical exfoliation. We present a non-destructive protocol for exfoliating crystals and show success to 30~$\mu$m, where sample quality is observed to improve with successive cleaving. Higher temperature magnetic features at 10~K/12~K are found to emerge through cleaving, both with and without induced sample damage. In both cases, we link these additional magnetic features to a persistence of C2/m structure to the low-temperature regime.
\end{abstract}

%
%
%
%
%

\section{Introduction}

The van der Waals layered transition-metal trihalide $\alpha$-RuCl$_3$ has generated much attention in recent years as a prime candidate material for realizing the elusive Kitaev quantum spin liquid (QSL). The Kitaev model, first proposed in 2006, is a rare example of an exactly solved model with a QSL ground state \cite{Kitaev2006}. In this model, spins are arranged on a honeycomb lattice, with highly anisotropic bond-directional Ising-type interactions between each neighbor \cite{Kitaev2006}. In this configuration, the single site is unable to satisfy all three neighboring spin configurations at once, resulting in exchange-driven magnetic frustration and an absence of long-range magnetic order found in the QSL phase. The resulting fractionalized excitations are proposed to have potential applications in fault-tolerant topological quantum computation, making the identification of candidate materials particularly compelling \cite{Aasen2020,Trebst2022,Hermanns2018,Motome2020}. 

In $\alpha$-RuCl$_3$, Ru$^{3+}$ ions arrange in a near ideal 2D honeycomb configuration, octahedrally coordinated by six Cl$^{-}$ ligand ions \cite{Matsuda2025,Plumb2014}. Layers are weakly bonded by van der Waals forces, readily cleaved through mechanical exfoliation down to the monolayer \cite{Yang2023,Massicotte2024,Weber2016,Zhou2019}. Its promise as a Kitaev QSL candidate stems from the local environment of the ruthenium ions, on-site Coulomb repulsion, and significant spin-orbit coupling, giving rise to highly anisotropic interactions via the Jackeli and Khaliullin mechanism \cite{jackeli}. 

In a true QSL ground state, quantum fluctuations prevent the stabilization of long-range magnetic order down to 0~K; however, candidate materials often magnetically order at sufficiently low temperatures due to the presence of non-Kitaev interactions. $\alpha$-RuCl$_3$ is no exception, adopting a well-established low-temperature zig-zag antiferromagnetic arrangement of Ru moments \cite{Sears2015,Johnson2015}. However, it is still considered a strong candidate for Kitaev physics, with signatures of fractionalized excitations reported in inelastic neutron scattering measurements \cite{Banerjee2016,Banerjee2017,Do2017}. Extensive works have gone towards an understanding of the magnetic Hamiltonian, and the determination of accompanying interaction strengths \cite{Matsuda2025,Suzuki2021,Sears2020}.

The structural, magnetic, electronic, and thermal transport properties of $\alpha$-RuCl$_3$ have been extensively investigated in the past decade or so \cite{Matsuda2025,Takagi2019}. Interest was further bolstered by the observation of a quantized thermal Hall effect in 2018, a promising indicator of exotic fractionalized excitations \cite{Kasahara2018}; however, difficulties in reproducing the result have slowed progress in understanding this material \cite{Yamashita2020,Czajka2021,Lefrancois2022,Bruin2022_nat}. Experimental difficulties are typically traced back to questions regarding the sample quality, mainly due to stacking defects typical of van der Waals layered materials \cite{Winter2017}.

\begin{figure}
    \centering
    \includegraphics[width=\linewidth]{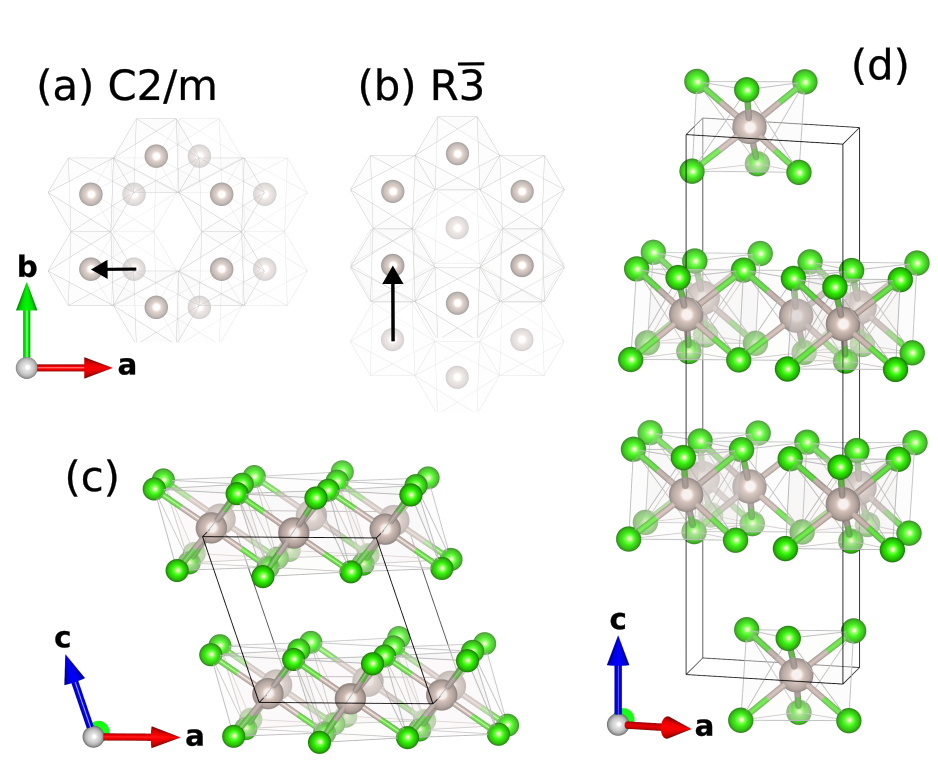}
    \caption{Crystal structures of $\alpha$-RuCl$_3$ including (c) high-temperature monoclinic (C2/m) structure and (d) low-temperature hexagonal (R$\bar{3}$) structure. The black arrows in (a) and (b) indicate the stacking sequence of neighbouring layers with \textit{a}-directional honeycomb stacking [-1/3,0,1] in the C2/m phase and \textit{b}-directional stacking [0,1/3,1] in the R$\bar{3}$ phase.}
    \label{fig:structure}
\end{figure}

There have been substantial efforts to understand the structure and magnetism of various $\alpha$-RuCl$_3$ crystals. It is widely accepted that ‘good’ samples are characterized by a single N\'eel ordering transition around 7~K, while so-called ‘bad’ crystals exhibit multiple magnetic transitions at higher temperatures (10~K, 12~K, 14~K) \cite{Kim_2025,Namba2024,Zhang2023,Zhang2024,Kim_APL,Kubota2015,Mi2021}. Neutron diffraction measurements have shown that the 7~K and 14~K features correspond to three-layer and two-layer zigzag magnetic ordering, respectively \cite{Banerjee2016,Balz2021}.

The ‘good’ samples tend to be larger, thick crystals obtained by sublimation growth that undergo a complete first-order structural transition at approximately 150~K \cite{Namba2024,Kim_APL,Yan2023,Wolf2022,May2020}. Determination of the room-temperature and low-temperature structures has been complicated by stacking faults, reported as P3$_1$12, C2/m, and R$\bar{3}$ in the literature, all differing primarily in stacking pattern \cite{Fletcher1967,Matsuda2025}. It is now largely accepted that high quality bulk $\alpha$-RuCl$_3$ crystals transition from a room-temperature monoclinic (C2/m) structure to a rhombohedral (R$\bar{3}$) structure \cite{Namba2024,Park2016,Sears2023,Mu2022}. As illustrated in Fig.~\ref{fig:structure}, in the C2/m structure, the next layer is shifted by 1/3 of \textit{a} lattice unit vector while the next R$\bar{3}$ layer is shifted by 1/3 of \textit{b} lattice unit vector. The temperature range of the structural thermal hysteresis is another clear marker of sample quality, with ‘good’ samples having a small hysteresis feature opening and closing within 20-30~K \cite{Namba2024,Mi2021}. 

The ‘bad’ crystals tend to be thin and small in size, exhibiting multiple magnetic transitions that coexist with the 7~K transition and a structural thermal hysteresis which spans a larger temperature range \cite{Kim_APL}. 

Several studies have investigated $\alpha$-RuCl$_3$ with thickness in the nanometer to monolayer range \cite{Du2019,Zhou2019,Mashhadi2018,Lee2021,Massicotte2024,Pai2021,Ziatdinov2016,Gronke2018}. In addition to the ‘good’ and ‘bad’ samples, some exfoliated flakes maintain C2/m structure to low temperatures and show a single 14~K magnetic transition \cite{Aristizabal2025}. This is consistent with the single 14~K transition observed in measurements of $\alpha$-RuCl$_3$ powder and polycrystalline samples \cite{Cao2016,Fletcher1967}. Importantly, the structural transition is severely broadened and sometimes even absent in very small crystals.

Efforts to characterize $\alpha$-RuCl$_3$ have focused heavily on the ‘good’ samples, while samples deemed of poorer quality, exhibiting higher transition temperatures, have received significantly less attention. These samples retain C2/m crystal symmetry at low temperature, and many aspects of their two-layer periodic magnetic order are not understood. For example, the origin of the higher magnetic transition temperature is unknown, and how the magnetic transition behaves when an in-plane magnetic field is applied is also not well established. 

The objectives of the current paper are twofold. The first is to investigate how crystals with a single magnetic transition at 7~K are related to those with a single magnetic transition at 14~K. Noting that thick and thin as-grown crystals are associated with these magnetic transitions, we decouple sample-dependence and thickness-dependence by reducing thickness through mechanical exfoliation and study their magnetic and structural properties \cite{Kim_APL}. The second, more practical question is whether it is possible to exfoliate from the surface without damaging the bulk crystal. In other words, we aim to find a handling protocol for cleaving this material.

\section{Method} 

Single crystal samples of $\alpha$-RuCl$_3$ were grown using the previously described chemical vapour transport method \cite{Kim_APL}. Thick crystals with large face size were screened and aligned at room temperature using a Rigaku Smartlab X-ray diffractometer with Cu K$_\alpha$ source. A total of 11 crystals of 1-2$^\circ$ mosaicity based on (004) Bragg peak were chosen for magnetic susceptibility measurement. We show results for only three of these samples (labeled S1, S2, S3) where careful exfoliation of a small thickness (approximately 10-20~$\mu$m) was achieved, and the bulk was left undamaged after the first exfoliation. Of the eight samples not shown, seven were damaged during cleaving and one was used to investigate the impact of temperature cycling. Temperature cycling was also investigated on S3 after being cleaved to 30~$\mu$m.

\begin{figure}
    \centering
    \includegraphics[width=\linewidth]{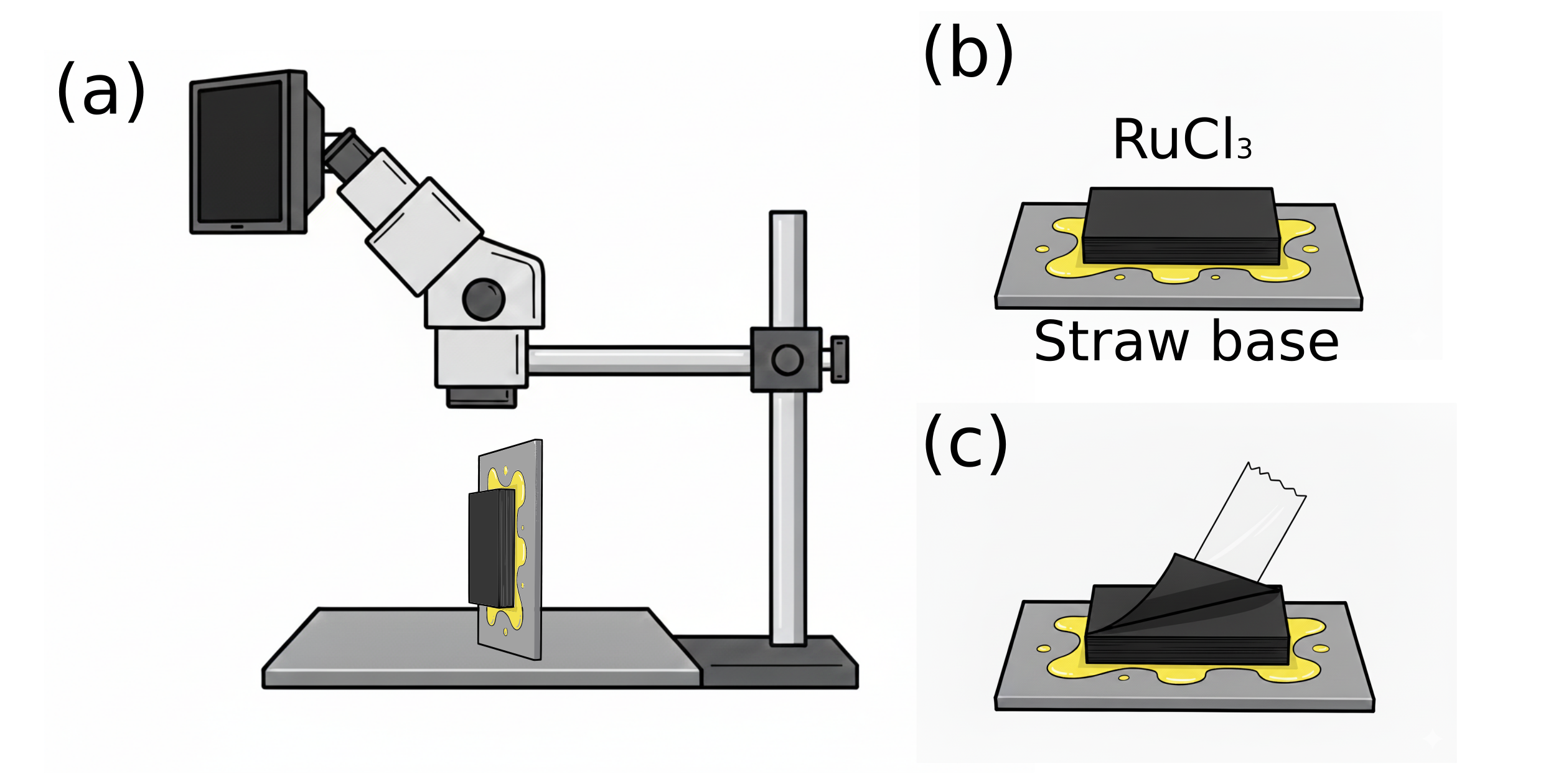}
    \caption{Schematic of the cleaving process between magnetic susceptibility measurements generated by Gemini (2.5 Flash). (b) $\alpha$-RuCl$_3$ samples (black) are mounted on a plastic straw (grey) using a GE varnish (yellow). (c) Samples are mechanically exfoliated using scotch tape, discarding the removed portion and continuing with what remains of the bulk. (a) After cleaving, samples are secured vertically, and thickness is estimated using a calibrated optical microscope.}
    \label{fig:cleaving}
\end{figure}

Samples were mounted with GE-varnish on a rectangular plastic base made from a flattened piece of translucent plastic straw (Fig.~\ref{fig:cleaving}b). In previous studies, in-plane directional dependence was observed in magnetic susceptibility in the C2/m phase due to an absence of C$_3$ symmetry \cite{Kim_structural,LampenKelley2018}. However, in the low-temperature R$\bar{3}$ phase, the in-plane directional dependence of magnetic susceptibility disappears, with the largest change in susceptibility from C2/m to R$\bar{3}$ observed with the field applied along the \textit{a}-direction \cite{Kim_structural}. To use temperature-dependent magnetic susceptibility as a proxy for the structural transition, and therefore to make the structural hysteresis feature most visible, crystals were aligned with the field directed in-plane along the \textit{a}-direction.

Measurements of magnetic susceptibility were conducted using a SQUID magnetometer (Quantum Design MPMS XL) under a static field through a temperature range of 3-300K. A magnetic field of 0.5T was applied in all but the thinnest sample (labeled S2 below), where the magnetic field from 0.5~T to 4~T was used to compensate for reduced sample volume as layers were removed. The measurement sequence consisted of a zero-field cool to 3K, and data collection through the full heat/cool range in an applied field, with a maximum cool/heat rate of 10K/min.

Between each magnetic susceptibility measurement, the mounted sample was removed from the MPMS and carefully cleaved using Scotch tape to peel back and discard layers (Fig.~\ref{fig:cleaving}c). Careful removal of layers from the sample corner was crucial in avoiding damage to the bulk. 

Sample thickness was primarily estimated using a stereoscopic microscope (Nikon SMZ745T) calibrated using a 0.01mm division microscope calibration slide. The mounted sample was positioned under the microscope camera on its side (Fig.~\ref{fig:cleaving}a), held in place by cross lock tweezers. With the edge of the sample in focus, an image measurement software was used to mark edges and determine the approximate thickness of the sample along the stacking direction (with an estimated uncertainty of $\sim$10~$\mu$m). Many thickness measurements on all sides were averaged in the reported thicknesses.

For S1, which is large enough for reliable mass measurements, thickness was inferred from sample mass and face area, using measurements before and after cleaving to approximate the layers removed. In this approach, the microscope was coarsely calibrated using mm-grid and the sample was laid flat under the microscope camera, where the area was estimated using the image measurement software. An average face area before and after cleaving was paired with the difference in mass, and using the density of $\alpha$-RuCl$_3$ (3.92~g/cm$^3$ based on the C2/m structure reported by Johnson et al.), the approximate thickness removed was calculated \cite{Johnson2015}. Using a direct microscope measurement of the final sample thickness, intermediate thicknesses could be deduced using the calculated thickness removed. It should be noted that the direct thickness measurement of S1 (140~$\mu$m) is only representative of the thickest region, with a small section measured to be $>$20~$\mu$m after the final exfoliation. Since S2 and S3 were too small for a reliable mass measurement, thickness was estimated entirely using direct microscope measurements.

In some cases, mechanical exfoliation resulted in a thickness change of less than 10~$\mu$m, making the thickness indistinguishable within our measurement precision before and after cleaving. Despite differences in the susceptibility data, the thickness is estimated to be the same. This includes thicknesses of 310~$\mu$m and 260~$\mu$m for S1 and 20~$\mu$m for S2. Magnetic susceptibility is still shown in these cases, but legend entries are doubled (Fig.~\ref{fig:data}).

\section{Results}

\begin{figure*}
    \centering
    \includegraphics[width=\linewidth]{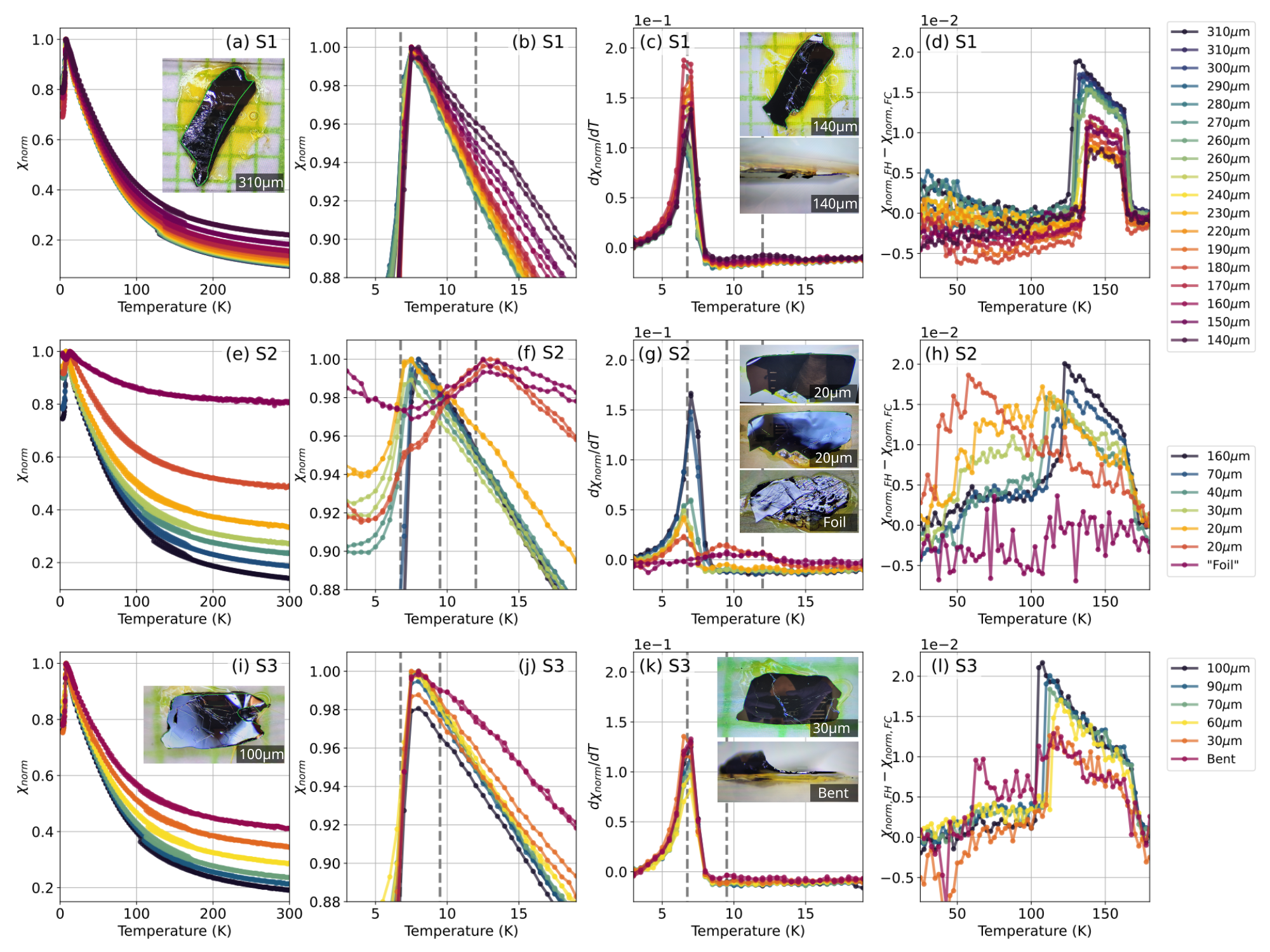}
    \caption{(a) Normalized magnetic susceptibility as a function of temperature for $\alpha$-RuCl$_3$ (a) S1, (e) S2, and (i) S3 through a series of mechanical exfoliations. Trials reported with the same thickness were cleaved, but the measurement change was below the uncertainty. Magnetic transition close-up shown for (b) S1, (f) S2, and (j) S3, marked with grey dashed lines at observed magnetic transitions (T = 6.75~K, 9.5~K, and 12~K). First temperature derivative of normalized magnetic susceptibility of (c) S1, (g) S2, and (k) S3 in the low-temperature range is plotted to show magnetic features more clearly, with dashed grey lines again marking observed transitions. The difference between field heating and cooling magnetic susceptibility data is shown for (d) S1, (h) S2, and (l) S3 over the temperature range capturing the structural transition. Microscope views of sample faces are shown in figure insets, with corresponding thickness labeled. Insets in (g) are shown from top to bottom in order of cleaving, corresponding to the yellow, orange, and purple data sets, respectively. See text for further details.}
    \label{fig:data}
\end{figure*}

Fig.~\ref{fig:data} presents magnetic susceptibility measurements for three single crystal samples of $\alpha$-RuCl$_3$ (S1, S2, and S3) as the thickness is altered through mechanical exfoliation. Comparison of these three samples illustrates the impact of structural disorder introduced by cleaving on magnetic properties.

Obtaining an accurate sample mass became challenging as volume was reduced, so susceptibility has been normalized to the maximum magnetization value ($\chi_{norm}$). Due to this normalization, background contribution becomes larger as thickness is reduced, manifesting as a gradually upward shifting signal in the susceptibility, especially visible in the high-temperature range (Fig.~\ref{fig:data}a,e,i). 

While the first column of panels (Fig.~\ref{fig:data}a,e,i) shows the overall temperature dependence of magnetic susceptibility, the middle two columns (Fig.~\ref{fig:data}b,c,f,g,j,k) show the data for the magnetic transition at low temperatures. All three samples initially show a single magnetic transition to zigzag long-range order at approximately 7~K, evident as a sharp cusp in the close-up plot of the low-temperature range shown in Fig.~\ref{fig:data}b,f,j. 

The last column (Fig.~\ref{fig:data}d,h,l) displays the differential susceptibility to emphasize thermal hysteresis associated with the structural transition. $\chi_{norm}(T)$ on heating is subtracted from $\chi_{norm}(T)$ on cooling. On heating, the transition from low-temperature R$\bar{3}$ structure to high-temperature C2/m consistently appears at approximately 170~K in S1, S2, and S3; however, on cooling, the temperature of transition from C2/m to R$\bar{3}$ is highly sample dependent and changes as the samples are mechanically exfoliated. 

Prior to exfoliation, the primary differences between samples S1 and S2 are in quality and thickness. S1 is initially 310~$\mu$m along the stacking direction, with a clear and sharp hysteresis in the range of 125-170~K as shown in Fig.~\ref{fig:data}d, while S2 is roughly half as thick with a stepped feature in Fig.~\ref{fig:data}(h) which opens at 170~K but does not fully close until 43~K. A structural hysteresis spanning a large temperature range indicates an incomplete structural transition, as has been previously shown with x-ray diffraction \cite{Zhang2024, Kim_2025}.

S3 is the thinnest of the three presented samples, with a starting thickness of 100~$\mu$m. Similar to S1, S3 exhibits a clear structural hysteresis feature, fully closing with no low-temperature portion as seen in S2 (Fig.~\ref{fig:data}l). However, the hysteresis temperature range is larger than in S1, opening around 170~K but not closing until 100~K.

As layers are removed from S1, the temperature range of the structural hysteresis loop shrinks, as is obvious in Fig.~\ref{fig:data}d. Through a thickness range of 310~$\mu$m to 230~$\mu$m, the portion of the hysteresis between 130~K and 135~K gradually becomes smaller, until fully closing at 135~K for a thickness of 220~$\mu$m. Further decrease in thickness does not change the hysteresis behaviour. Magnetic properties are also affected as the thickness of S1 is decreased with successive mechanical exfoliation. A broad feature emerges at approximately 12~K for a thickness of 160~$\mu$m, becoming more pronounced as the thickness is further reduced to 140~$\mu$m (marked with a dashed grey line in Fig.~\ref{fig:data}). The appearance of this hump is already visible in the close-up of the magnetic transition (Fig.~\ref{fig:data}b), but d$\chi_{norm}$/dT is plotted in Fig.~\ref{fig:data}c for the same temperature range to show these features more clearly. Although a small thermal hysteresis is also observed in the low-temperature magnetic susceptibility on heating/cooling, this is in the opposite direction when compared to the structural hysteresis and does not appear to depend on mechanical exfoliation. 

The exfoliation progression of S1 should be contrasted with that of S2 shown in Fig.~\ref{fig:data}f, which not only develops a broad feature at 12K at a thickness of 30~$\mu$m, but also a feature at 9.5K. This is most visible in the orange curve for a thickness of 20~$\mu$m. We observe that the 7~K transition completely disappears in the thinnest sample data, labelled ``Foil'', due to our inability to accurately assign a thickness ($<$10~$\mu$m). As before, d$\chi_{norm}$/dT is plotted in Fig.~\ref{fig:data}g to more clearly distinguish magnetic features, with 9.5~K and 12~K marked as dashed grey lines. Insets display the sample face of S2 at a thickness of 20~$\mu$m and after the final exfoliation (corresponding to the yellow, orange, and purple lines). The sample damage becomes more apparent as layers are removed, observable in the increasingly crinkled sample surface.

The structural thermal hysteresis behaviour in S2 with successive layer removal is also markedly different than that of S1. Plotted in Fig.~\ref{fig:data}h, the hysteresis remains partially open until approximately 43~K even before cleaving (dark blue curve). The hysteresis sliver in the range of 40-115~K grows as layers are removed, appearing as a shifting of weight from the 115-170~K range, clearest in the orange curve. In the final exfoliation of S2 the small sample mass diminishes the signal, but it appears the structural hysteresis vanishes entirely (purple curve in Fig.~\ref{fig:data}h).

In S2, the emergence of additional magnetic transitions with sample damage is consistent with previous studies reporting the appearance of multiple magnetic transitions with deformation\cite{Cao2016}, and the known two-layer periodicity of 12~K/14~K magnetic transitions is explainable using the C2/m structure if the high-temperature phase persists to low temperatures \cite{Johnson2015}.

The puzzling observation for S1, which exhibits additional magnetic transitions emerging at 12~K despite improved thermal hysteresis behaviour, is not universal. Similar to S1, careful mechanical exfoliation resulted in a shrinking structural hysteresis from an approximate starting thickness of 100~$\mu$m to 60~$\mu$m (Fig.~\ref{fig:data}l). Further cleaving to 30~$\mu$m extended the hysteresis by a single temperature step size (2.5K, visible in Fig.~\ref{fig:data}l); however, additional higher temperature magnetic transitions were not observed (Fig.~\ref{fig:data}k). In efforts to further reduce the thickness, the sample was bent (inset of Fig.~\ref{fig:data}k) and, as expected, the structural hysteresis extended to lower temperatures and a second broad magnetic transition appeared.

Therefore, 30~$\mu$m is the thinnest sample for which we observed a clear structural transition from high-temperature C2/m to low-temperature R$\bar{3}$ without the appearance of high-temperature magnetic transitions.

\section{Discussion}
\subsection{Modeling}

\begin{figure}
    \centering
    \includegraphics[width=\linewidth]{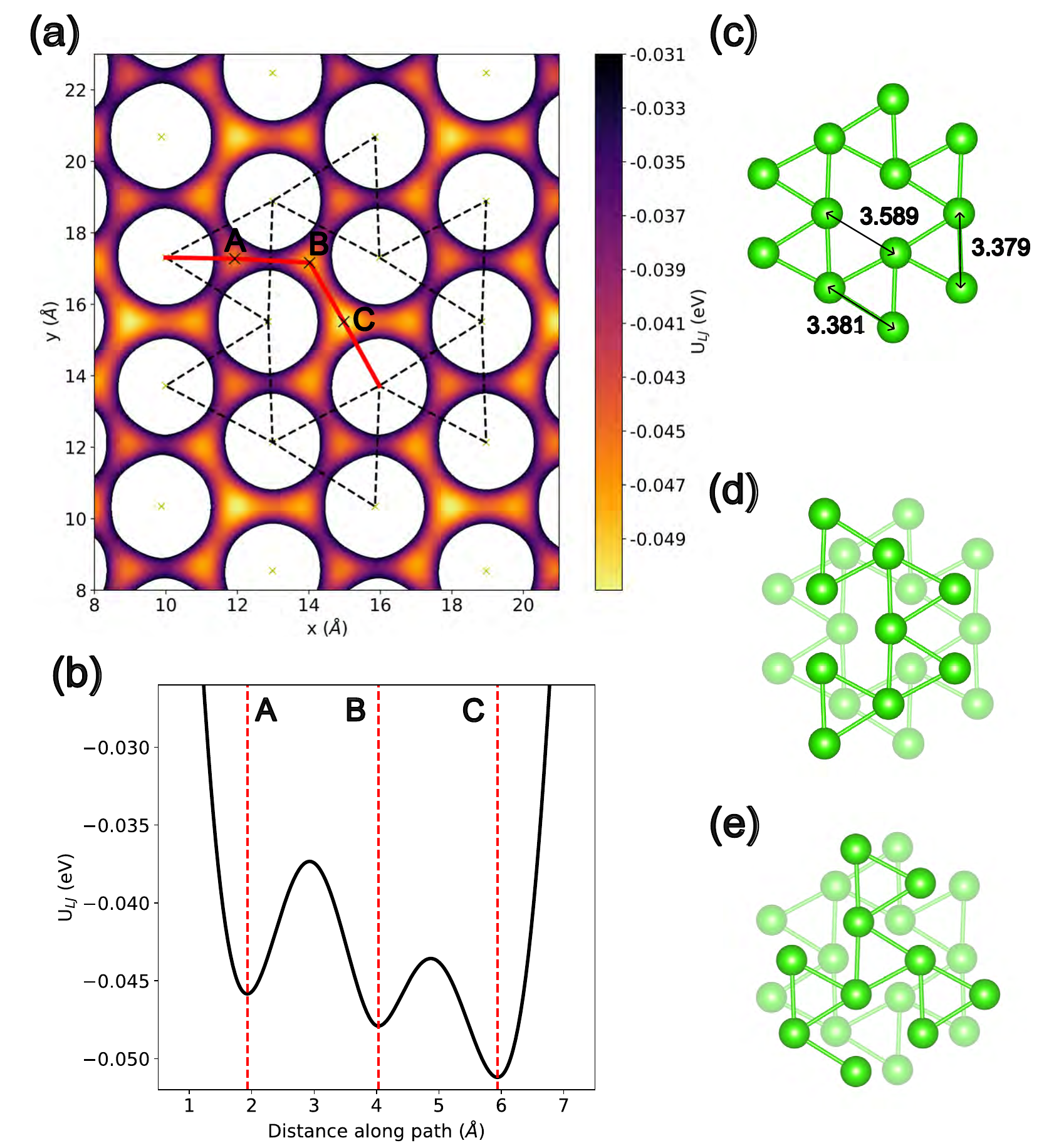}
    \caption{(a) Pseudocolor contour plot of Lennard-Jones electrostatic potential 3\AA{} above the triangular Cl layer, with yellow 'x's marking Cl$^-$ positions and the underlying lattice shown with dashed black lines. Possible positions of Cl ions in the next layer are marked as positions A/B/C, with the A/C positions occupied in the C2/m structure and the B positions occupied in the R$\bar{3}$ structure. Solid red line traces the line cut shown in (b), where A/B/C positions are marked with red dashed lines. Triangular Cl lattice with labeled bond lengths is shown in (c), made more transparent in (d)/(e) where the overlaid Cl lattice corresponds to C2/m structure (d) or R$\bar{3}$ structure (e).}
    \label{fig:model}
\end{figure}

Since the structural transition between the room temperature C2/m structure and the low-temperature R$\bar{3}$ structure seems to be at the heart of the sample dependence, we carried out a simple modeling to understand the physics of this transition. It is important to recognize that the difference between the two structures comes from the different stacking sequence of RuCl$_3$ layers, as illustrated in Fig.~\ref{fig:structure}a. Each RuCl$_3$ layer is, in fact, made up of three layers of ions: a honeycomb layer of Ru ions sandwiched by chlorine layers forming a triangular lattice, which means that the stacking sequence can be described as Cl-Ru-Cl-Cl-Ru-Cl-... Therefore, the crucial difference between C2/m and R$\bar{3}$ boils down to the way the two chlorine triangular lattices stack on top of each other. Another important detail of the Cl layer is that the bond lengths between Cl ions surrounding Ru ions (i.e., part of RuCl$_6$ octahedra) are shorter than those surrounding the void in the Ru honeycomb lattice, as shown in Fig.~\ref{fig:model}c. The bond length difference between Cl ions gives rise to an electrostatic potential energy, which is inequivalent for the three possible Cl positions in the next layer.

We carried out a simple modeling of the potential energy landscape arising from the Cl layer using the experimental bond lengths shown in Fig.~\ref{fig:model}c. We assume van der Waals interaction and use the Lennard-Jones potential with the parameter given for argon (same valence electron configuration as Cl$^-$) \cite{a&m}. The potential at 3\AA{} above the Cl-layer is plotted in a pseudocolor contour plot in Fig.~\ref{fig:model}a. The underlying Cl lattice is represented as thin dashed lines, and the white circles are centered at the chlorine ion positions. As expected, there are three types of potential energy minima, indicated by A, B, and C. C is at the center of the void in the Ru honeycomb and therefore has the lowest energy. The potential energy along the red solid path is plotted in Fig.~\ref{fig:model}b, which clearly shows the different potential energy minima located at A, B, and C. When the next layer Cl occupies A and C, the structure becomes C2/m (Fig.~\ref{fig:model}d), while the crystal becomes R$\bar{3}$ when Cl occupies B sites (Fig.~\ref{fig:model}e). 

This simple modeling sheds light on the nature of the structural transition. Since the potential energy $U_A>U_B>U_C$, the energy difference between $2U_A+U_C$ and $3U_B$ will depend sensitively on structural details. While C2/m is preferred at room temperature, the energetics appear to change with lattice contraction, and R$\bar{3}$ seems to have a lower energy at low temperatures. Indeed, stacking pattern has been shown to depend sensitively on lattice changes via applied pressure as well \cite{Li2019,Stahl2024,Cui2017}, where application of hydrostatic pressure has been seen to shift the structural transition to higher temperatures \cite{Wolf2022}. It is then also plausible to expect that the energy balance is strongly affected by local strain, which could be generated by surface strain, temperature gradient, or even gluing. Some of these effects could provide likely explanations for our observations as discussed below.

\subsection{Data Interpretation}

We present several interesting observations from our study that warrant further discussion. The first is that we were able to reduce the width of thermal hysteresis by cleaving in both S1 and S3, indicating that the crystal quality could actually be improved by cleaving up to a point. The second is the appearance of the `undesirable' 12~K magnetic transition, despite improved crystal quality in S1. The third is that we were able to produce a 30~$\mu$m crystal via cleaving without introducing additional disorder for S3.

Since the range of the hysteresis is typically an indicator of crystal quality in $\alpha$-RuCl$_3$ \cite{Namba2024}, the gradual narrowing of the hysteresis in S1 through cleaving (fig~\ref{fig:data}d) could be due to a gradual reduction of sample defects as layers are removed from the surface. It should also be noted that after several initial cleavings, the hysteresis temperature range remains relatively stable as the thickness is further reduced beyond 210~$\mu$m. 
One possible explanation is a non-uniform distribution of defects, more densely populating the first 100~$\mu$m from the surface. 
However, the appearance of a magnetic feature at 12~K in S1 seemingly contradicts a gradual improvement in sample quality, with the presence of multiple magnetic transitions consistently being an indicator of poor sample quality in this study and previous works \cite{Namba2024, Kim_APL}.

The appearance of two-layer magnetic stacking (manifesting as magnetic transitions at 10~K/12~K/14~K) is not yet fully understood, but has been clearly shown to depend on sample damage and size. High-quality crystals with a narrow structural hysteresis and a single magnetic transition at 7~K develop 10~K/14~K anomalies after being cut. They also develop a diffuse x-ray scattering signal along the L-direction, similar to that observed in thin as-grown samples with many stacking faults \cite{Namba2024, Kim_APL}. In the extreme regime of deformation and damage of single crystals, the 7~K magnetic transition is fully suppressed, and there is only a single magnetic transition at 14~K \cite{Cao2016}. Our measurements for S2 are consistent with these observations, where the 7~K magnetic transition and structural hysteresis are completely suppressed in the final magnetic susceptibility measurements at its most damaged and thinnest (Fig.~\ref{fig:data}f/h, purple curves).

In samples dominated by the 14~K transition, magnetic stacking is described using the high-temperature C2/m structure \cite{Johnson2015}. Recent classical Monte Carlo simulations support the connection between the two-layer magnetic stacking and C2/m structure, where it has been demonstrated that differing honeycomb bond lengths enhance the N\'eel temperature from the R$\bar{3}$ $\sim$7.5~K to a C2/m $\sim$12.5~K \cite{Cen2025}. This persistence of C2/m structure in the low temperature region is fully consistent with our study, where we observe damage-induced widening of the structural hysteresis accompanied by the emergence of additional magnetic transitions.

A compelling explanation for the higher temperature magnetic feature in magnetic susceptibility measurements of S1 is the presence of a critical thickness for structural transition. Although the feature could be interpreted as damage, this conflicts with the unchanged narrow structural hysteresis. As demonstrated in S2, sample damage is evident in a widening of the structural hysteresis (fig ~\ref{fig:data}h); however, in S1 the 12~K magnetic feature appears while the hysteresis range remains unchanged. This behaviour in S1 suggests the C2/m structure could persist into the low-temperature regime without inducing damage (Fig.~\ref{fig:data}b). Due to the crystal shape, S1 had a thin ledge of $\sim$10~$\mu$m in the final measurement, and if this portion of the sample were unable to transition into R$\bar{3}$ due to insufficient thickness/size, this would be undetectable in the hysteresis but visible as a two-layer stacked magnetic ordering. This is further supported by the slight reduction of hysteresis height between 150~$\mu$m and 140~$\mu$m, where persistence of C2/m would reduce the bulk response of the transition to R$\bar{3}$. Combined with results from S3 showing a complete structural transition and a single 7~K magnetic transition at an approximate thickness of $\sim$30~$\mu$m, it is possible that there is a critical thickness in the range of $\sim$10-30~$\mu$m, below which RuCl$_3$ maintains the C2/m structure through the full temperature range. One possible reason for this might be surface strain that causes C2/m structure to be more stable near the surface region of a thick crystal or the entirety of a thin crystal.

Dependence of the structural transition on thickness or size can also be reconciled with these studies on damaged and polycrystalline samples. Damage-induced stacking faults may create domains of reduced volume, effectively reducing crystal size in a fraction of the bulk and preventing a complete structural transition. This is consistent with the observations that small crystals often exhibit only a single 14~K magnetic transition \cite{FroudePowers2024}, few-layer flakes maintain C2/m symmetry with a single 14~K magnetic transition \cite{Massicotte2024, Pai2021}, powders consistently have only a single 14~K magnetic transition \cite{Cao2016,Fletcher1967}, and the 7~K magnetic transition can be fully suppressed with sufficient manual deformation.

It should also be noted that the introduction of stacking faults through temperature cycling is inevitable, and has been thoroughly studied \cite{Sears2023}. Some fraction of the crystal is consistently seen to maintain C2/m structure (expressing all three domains) well into the low-temperature phase, even in high-quality single crystals \cite{Sears2023,Kim_structural}. If the C2/m structure is maintained through the full temperature range (3-300~K), there would be little evidence of this in the hysteresis loop observed in magnetic susceptibility (though a new magnetic transition would appear). However, if induced stacking faults are minimal enough to result in a fraction of the crystal which eventually transitions into the low-temperature phase at a lower temperature, this would appear as a long sliver-like tail on the main hysteresis feature.

To investigate the impact of temperature cycling on our study, we repeatedly measured the magnetic susceptibility of one thick ($\sim$350~$\mu$m) and one thin ($\sim$30~$\mu$m) sample, both exhibiting a single magnetic transition at approximately 7~K. In the thick sample, defects induced via temperature cycling and adhesive are not discernible in either the structural hysteresis loop or magnetic transition. Interestingly, in the thin sample, temperature cycling appears to cause a slight narrowing of the main structural hysteresis loop; however, the emergence of a small sliver suggests part of the sample is becoming more damaged and undergoing an incomplete structural transition, as expected for samples with increased stacking faults (see Supplementary Material). The requirement of an adhesive, therefore, limits this study, and future probing of thinner samples must be completed without this additional damage-inducing factor.

\section{Conclusions}
We have presented a magnetic susceptibility study of the thickness-dependent behaviour of magnetic and structural transitions in $\alpha$-RuCl$_3$. To this end, we have demonstrated $\alpha$-RuCl$_3$ can be cleaved in a non-destructive manner through careful mechanical exfoliation. Without inducing sample damage, a minimum thickness of $\sim$30~$\mu$m was achieved.

In successful mechanical exfoliation, we observe a narrowing of structural hysteresis with successive cleaves, suggesting that sample quality can be gradually improved by removing surface layers. In samples damaged by unsuccessful mechanical exfoliation, we consistently observe a widening of the structural hysteresis with successive cleaves, accompanied by the appearance of 10~K and 12~K magnetic transitions. The behaviour is consistent with previous specific heat measurements of deformed samples, and can be explained by the damage-induced persistence of C2/m to the low-temperature regime \cite{Cao2016}. We argue our results are consistent with a critical thickness required to undergo the structural transition to low-temperature R$\bar{3}$ phase, and provide a thickness range of $\sim$10-30~$\mu$m based on the appearance of the 12~K magnetic transition without sample damage.

\section*{Acknowledgement}
Work at the University of Toronto was supported by the Natural Sciences and Engineering Research Council (NSERC) of Canada through the Discovery Grant No. RGPIN-2025-06514, and by the Canada Foundation for Innovation (CFI) and the Government of Ontario for Project No. 36404.

\bibliography{iopart-num}

\end{document}